\documentstyle[twoside,fleqn,espcrc2,epsfig]{article}

\newcommand{\be}{\begin{eqnarray}}
\newcommand{\ee}{\end{eqnarray}}
\newcommand{\ben}{\begin{eqnarray*}}
\newcommand{\een}{\end{eqnarray*}}

\def\lab#1      {\hbox{\small #1} }
\def\diffn#1	  {\Delta^{-}_{#1}}
\def\mb#1       {\mbox{\boldmath $#1$}}
\newcommand{\AmS}{{\protect\the\textfont2
  A\kern-.1667em\lower.5ex\hbox{M}\kern-.125emS}}

\title{Vortices in $SO(3)\times Z(2)$ simulations}
\author{Andrei Alexandru and Richard W. Haymaker\thanks{Presented 
        by A. Alexandru at Lattice 2000, July 17 - 22, Bangalore.
   Work partially supported by the U. S. Department of Energy under grant 
               DE-FG05-91 ER 40617.}
               \address{Dept. of Physics and Astronomy, 
       	Louisiana State University, Baton Rouge, Louisiana,
       	70803  USA}        }
\begin{document}

\begin{abstract}
We explore simulations on periodic lattices in the Tomboulis $SO(3)\times Z(2)$ formulation. The dynamical variables are constrained. We propose an update algorithm that satisfies the constraints and is straightforward to implement. We show how boundary conditions put constraints on the configuration space.
\end{abstract}

\maketitle

\section{Introduction}

The center vortices seem to play an important role in disordering the Wislon loop. The mechanism is explained in a number of papers using an intuitive idea about vortices. We decided to investigate the vortices using Tomboulis formalism [1,2]. The Tomboulis formalism is based on spliting the $SU(N)$ degrees of freedom into $SU(N)/Z(N)$ variables living on links and $Z(N)$ variables living on plaquettes. The new variables are constrained and for the free boundary case the constraint amounts to a coincidence between the thick and thin monopoles.

In trying to simulate the new variables we have to solve two problems. The first problem regards the boundary conditions: the Tomboulis analysis was done for a free boundary lattice. We need periodic boundary condition for our simulation and we have to understand how they influence the constraint on the variables. The second problem is to find a suitable definition for the constraining set. The definition that Tomboulis derived was rather awkward to implement in a numerical investigation.
 
\section{Derivation}

From now on we are going to deal only with $SU(2)$. We start with the Wilson partition function for $SU(2)$:
$$
Z=\int [dU] e^{\frac{\beta}{2}\sum_p Tr(U_p)}.
$$
Following Tomboulis we re-write it:
$$
Z=\int [dU]' \sum_{\sigma} C[\sigma\eta] e^{\frac{\beta}{2}\sum_p |Tr(U_p)| \sigma(p)},
$$
where
$$
C[\alpha]=\sum_{\tau\in{\cal A}} \prod_p \chi_{\tau(p)}(\alpha(p)) \prod_b \delta(\tau(\hat{\partial}b)).
$$ 
The set:
$$
{\cal A}=\{\alpha:P\rightarrow Z(2)\}
$$
 is the set of all $Z(2)$ configurations defined on plaquettes and $\eta(p)=
 \lab{sgn} Tr(U_p)$ is the sign of the plaquette. We see that the function $C$ is defined as a summation over all configuration $\tau$ that obey the constraint $ \prod_b \delta(\tau(\hat{\partial}b))=1$ of the function $\prod_p \chi_{\tau(p)}(\alpha(p))$ where: 
\begin{eqnarray*}
\chi_1(\pm 1)&=&1, \\
\chi_{-1}(\pm 1)&=&\pm 1,
\end{eqnarray*}
are the characters of the two irreducible representations of $Z(2)$.

The partition function is then given by an integration over the $SO(3)$ degrees of freedom that live on the links and a summation over the $Z(2)$ degrees of freedom that live on plaquettes. The integration over the $SO(3)$ is unconstrained. The summation over the $Z(2)$ variables seems to be unconstrained too but the $C$ function will act as a constraint. For example for free boundary conditions the $C$ function has the form:
$$
C[\alpha]=\prod_c \delta(\alpha(\partial c)).
$$

To make these things clear we will introduce some notation.

\section{Notation}

For any two configurations $\alpha,\beta\in {\cal A}$ we define:
$$
\langle \alpha,\beta \rangle =\prod_p \chi_{\alpha(p)} (\beta(p)).
$$
This bracket is giving some degree of superposition between the two configurations. We see that $\chi_{a}(b)=-1$ only when both $a$ and $b$ are $-1$. Thus the bracket is going to be $1$ if we have an even number of plaquettes that have both $\alpha(p)=-1$ and $\beta(p)=-1$ and is going to be $-1$ if we have an odd number of such plaquettes.

Here we list some properties of the bracket: 
\begin{eqnarray*}
\langle \alpha, \beta \rangle &=& \langle \beta,\alpha \rangle, \\
\langle \alpha, \beta\gamma\rangle &=& \langle\alpha,\beta \rangle \langle \alpha,\gamma\rangle, \\
\langle\alpha,{\bf 1}\rangle&=&1.
\end{eqnarray*}
where ${\bf 1}$ is the configuration with ${\bf 1}(p)=1$ for all plaquettes $p$.

In the definition of $C$ we have a summation over all configurations that have the property $\prod_b \delta(\tau(\hat{\partial}b))=1$. Define:
$$
{\cal C}=\{\tau\in {\cal A}|\prod_p \delta(\tau(\hat{\partial}b))=1\},
$$
the set of all such configurations. One important thing to note about the set ${\cal C}$ is that it forms a group under the multiplication law:
$$
(\alpha\beta)(p)=\alpha(p)\beta(p).
$$

Now for any subgroup ${\cal K}$  included in ${\cal A}$ ($\cal C$ is such a subgroup) define:
$$
\bar{\cal K}=\{\alpha\in {\cal A}|<\alpha,\beta>=1\,\,\, \forall \beta\in {\cal K}\}.
$$
In other words the set $\bar{\cal K}$ is made up from the elements that have the bracket with all the elements in the original set ${\cal K}$ equal with $1$. This is some sort of dual set since $\bar{\bar{\cal K}}={\cal K}$ and $|{\cal K}|\times|\bar{\cal K}|=|{\cal A}|$ (we denote with $|{\cal K}|$ the number of elements in the set ${\cal K}$).

Using this notation we will show that the ${\cal C}$ is just some sort of delta function and thus acting as a constraint.

\section{The $C$ function}

Using our notations  we write:
$$
C[\alpha]=\sum_{\tau\in {\cal C}} \langle \alpha,\tau\rangle.
$$
Now we remember that ${\cal C}$ is a group and using group summation invariance we can write:
$$
C[\alpha]=\sum_{\tau\in {\cal C}}\langle \alpha,\tau\tau_0\rangle=\langle\alpha,\tau_0\rangle\sum_{\tau\in {\cal C}}\langle \alpha,\tau\rangle=\langle\alpha,\tau_0\rangle C[\alpha].
$$
This is true for any $\tau_0\in{\cal C}$. Thus if we have at least one $\tau_0\in{\cal C}$ with $\langle \alpha, \tau_0 \rangle=-1$ then $C[\alpha]=-C[\alpha]$ and thus $C[\alpha]=0$. 

Now if we don't have any element in ${\cal C}$ that has the bracket with $\alpha$ equal to  $-1$ that means that $\langle\alpha,\tau\rangle =1$ for all $\tau\in{\cal C}$ and thus $\alpha$ is a element of the  dual set $\bar{\cal C}$. Moreover:
$$
C[\alpha]=\sum_{\tau\in {\cal C}}\langle\alpha, \tau\rangle =\sum_{\tau\in {\cal C}}1=|{\cal C}|.
$$

Summing up, we have:
$$
C[\alpha]=\left\{ 
\begin{array}{cc}
|{\cal C}|&\alpha\in\bar{\cal C}, \\
0&\alpha\not\in\bar{\cal C},
\end{array}
\right.
$$
and we see that the function $C$ is nothing more than the characteristic function for the set $\bar{\cal C}$ (up to a constant). Thus the role of the $C$ function in the partition function is to constrain the summation over the $Z(2)$ degrees of freedom to the set $\bar{\cal C}$.

\section{$\bar{\cal C}$ set properties}

In the free boundary conditions case Tomboulis[1] showed that any configuration that obeys the cubic constraint $\prod_c \delta(\sigma(\partial c))=1$ is allowed. That is a configuration $\alpha$ is allowed if and only if for any cube $c$ in the lattice we have an even number of plaquettes $p$ in the boundary of the cube that have $\alpha(p)=-1$ (i.e. $\alpha(\partial c)=1$). Now we investigate the allowed configuration set in the case of periodic boundary conditions.

In the periodic boundary conditions case all configurations $\alpha \in\bar{\cal C}$ obey the cubic constraint. To prove it assume that :
$$
\alpha(\partial c)=-1
$$
for some particular cube $c$. Then we take $\tau\in{\cal C}$ defined as:
$$
\tau(p)=\left\{
\begin{array}{cc}
-1&p\in\partial c, \\
1&p\not\in\partial c.
\end{array}
\right.
$$
Then $\langle\tau,\alpha\rangle=-1$ since there are an odd number of plaquettes that have both $\alpha=-1$ and $\tau=-1$ (this is due to the fact that all plaquettes on the faces of the cube have $\tau=-1$ but only an odd number of those plaquettes have $\alpha=-1$ since $\alpha(\partial c)=-1$). Thus we proved that if the configuration $\alpha$ doesn't obey the cubic constraint it is not a member of the $\bar{\cal C}$ set. 

We see that even in the periodic boundary conditions case the allowed configurations have to obey the cubic constrained. However, this is only a necessary condition and is not sufficient as for free boundary conditions. To prove this take the configuration:
$$
\alpha(p)=\left\{
\begin{array}{cc}
-1&p\in S_{12}, \\
1&p\not\in S_{12},
\end{array}
\right.
$$
that obeys the cubic constraint. Now choose:
$$
\tau(p)=\left\{
\begin{array}{cc}
-1&p\in P_{12}, \\
1&p\not\in P_{12},
\end{array}
\right.
$$
where $S_{12}$ is a co-plane  and $P_{12}$ is a  plane in the $12$ direction. Now it is easy to see that $\tau\in{\cal C}$ and $\langle\tau,\alpha\rangle=-1$ since $S_{12}$ and $P_{12}$ have only one plaquette in common. This means that although the $\alpha$ configuration obeys the cubic constraint it is not a member of the $\bar{\cal C}$ set.

Summing up these two observations we see that contrary to the free boundary conditions case the set $\bar{\cal C}$ (the set of allowed configurations) is only a subset of the set of all configurations that obey the cubic constraint. Thus the periodic boundary conditions impose further constraints on the allowed configurations set.

\section{Alternative definition for the set $\bar{\cal C}$}

The partition function (up to a constant) is:
$$
Z=\int [dU] \sum_{\sigma\eta\in\bar{\cal C}} e^{\frac{\beta}{2}\sum_p |Tr(U_p)|\sigma(p)}.
$$
We see that the set $\bar{\cal C}$ has all the informations regarding the  constraint. However, the definition that we have for the set $\bar{\cal C}$ is not suitable for numerical simulations. The set $\bar{\cal C}$ is defined in terms of the set ${\cal C}$ which in turn is defined using a constraint. Even in the case of  free boundary conditions where the set $\bar{\cal C}$ is given by the cubic constraint the definition is not easy to implement numerically. This is the reason we need a different definition for the set $\bar{\cal C}$.

To do this we employ the {\em star transformation} around a link $b$ defined as:
$$
\alpha_{\hat{\partial} b}(p)=\left\{
\begin{array}{cc}
-1&p\in\hat{\partial}b, \\
1&p\not\in\hat{\partial}b.
\end{array}
\right.
$$
These are $Z(2)$ configurations that are members of the set $\bar{\cal C}$. Since the set $\bar{\cal C}$ forms a group any product of such configurations is a member of the set. Now we define the set of all star transformations:
$$
{\cal D}=\{\alpha\in{\cal A}|\alpha=\prod_i\alpha_{\hat{\partial}b_i}\},
$$
which is obviously a subset of the set $\bar{\cal C}$. Now, the set ${\cal D}$ forms a group and we can generate it by starting with the identity (all plaquettes equal with $1$) and then doing star transformations around various links $b$ (i.e. flipping the signs of all six plaquettes around a link at once). If we employ this transformations for our update we are guaranteed to stay in the set $\bar{\cal C}$ and thus obeying the constraint. Our only problem is to determine whether or not we are sweeping through the whole set $\bar{\cal C}$. We were able to prove [3] that ${\cal D}=\bar{\cal C}$ and thus we have proved that by doing the star transformations we are covering the entire set $\bar{\cal C}$.
\vskip 1cm
We wish to thank E.T.Tomboulis and P. DeForcrand for helpful discussions.

\end{document}